\documentclass[manuscript]{acmart}

\AtBeginDocument{%
  }

\usepackage{multirow} 
\newtheorem{definition}{Definition}
\usepackage{amsmath} 
\usepackage{algorithm}
\usepackage{algorithmic}

\begin{document}

\title{Non-Progressive Influence Maximization in Dynamic Social Networks}


\author{Yunming Hui}
\affiliation{%
  \institution{University of Amsterdam}
  \city{Amsterdam}
  \country{The Netherlands}}
\email{y.hui@uva.nl}

\author{Shihan Wang}
\affiliation{%
  \institution{Utrecht University}
  \city{Utrecht}
  \country{The Netherlands}}
\email{s.wang2@uu.nl}

\author{Melisachew Wudage Chekol}
\affiliation{%
  \institution{Utrecht University}
  \city{Utrecht}
  \country{The Netherlands}}
\email{m.w.chekol@uu.nl}

\author{Stevan Rudinac}
\affiliation{%
  \institution{University of Amsterdam}
  \city{Amsterdam}
  \country{The Netherlands}}
\email{s.rudinac@uva.nl}

\author{Inez Maria Zwetsloot}
\affiliation{%
  \institution{University of Amsterdam}
  \city{Amsterdam}
  \country{The Netherlands}}
\email{i.m.zwetsloot@uva.nl}


\begin{abstract}

The influence maximization (IM) problem involves identifying a set of key individuals in a social network who can maximize the spread of influence through their network connections. With the advent of geometric deep learning on graphs, great progress has been made towards better solutions for the IM problem. In this paper, we focus on the dynamic non-progressive IM problem, which considers the dynamic nature of real-world social networks and the special case where the influence diffusion is non-progressive, i.e., nodes can be activated multiple times.
We first extend an existing diffusion model to capture the non-progressive influence propagation in dynamic social networks. 
%Furthermore, we propose the method, 
We then propose the method, DNIMRL, which employs deep reinforcement learning and dynamic graph embedding to solve the dynamic non-progressive IM problem. 
%To enable efficient processing of dynamic social networks, we modify an existing dynamic graph embedding method to capture the temporal changes of users which is crucial for solving dynamic IM problems. Then 
In particular, we propose a novel algorithm that effectively leverages graph embedding to capture the temporal changes of dynamic networks and seamlessly integrates with deep reinforcement learning. The experiments, on different types of real-world social network datasets, demonstrate that our method outperforms state-of-the-art baselines. 
% except on densely connected dynamic social networks.
\end{abstract}





\maketitle

\section{Introduction} \label{section_intro}
Influence maximization (IM) is a fundamental concept in the domain of social network analysis and is pivotal in applications ranging from viral marketing to idea propagation within social networks \cite{li2018influence}. At its core, IM seeks to identify a small and fixed-size subset of nodes (seed set) within a social network to maximize the spread of influence, e.g. opinions, behaviours or products.

In the IM problem, nodes in a social network have two states: active or inactive. Seed nodes are initially activated. Influence spreads through edges, allowing active nodes to activate connected inactive ones. The influence of the seed set is measured by the number of activated nodes or the duration all nodes remain active. For instance, in social media marketing, seed nodes are advertisers or paid opinion leaders to promote products. An active node endorses a product to contacts, while an inactive node has not yet accepted the produce.  Activation means persuading a user to accept and promote a product. 

IM problem has many variants and can be classified from multiple perspectives. The categorisation is mostly related to how the influence diffusion is regulated. In this paper, we concentrate on two aspects: static or dynamic, progressive or non-progressive. 
The classification of IM problems as static or dynamic hinges on the variability of the network structure. In static IM problems, the connections between nodes remain constant. In contrast, the connections evolve which makes influence paths change overtime. Additionally, IM problems can be categorized as progressive or non-progressive based on node state permanence. Progressive models keep nodes permanently active once activated, whereas non-progressive models allow nodes to deactivate and potentially reactivate \cite{fazli2014non,lou2014modeling}.

The classic and most-studied IM problem is static and progressive~\cite{banerjee2020survey}. However, non-progressive offers a realistic portrayal of scenarios where individuals’ attitudes or the credibility of information changes over time \cite{golnari2014revisiting, lou2014modeling}. An illustrative example is the gaming software industry, where users' engagement may fluctuate. Users might lapse into inactivity and require reactivation with the introduction of new features, necessitating targeted promotional efforts to regain their engagement. 
%Despite the inherent dynamics of networks, many studies use static diffusion models to simplify the IM problem, which can fail to capture the fluidity of influence diffusion. For example, static models might portray a user’s influence as unchanging, neglecting shifts in follower interests. 
Hence, in this paper, we focus on the dynamic non-progressive IM problem.

The classic IM problem, as a combinatorial optimization (CO) problem, is NP-hard \cite{kempe2003maximizing,li2023survey}. This underscores the exponential growth in the computational complexity required to find the optimal solution with the increased network size. In dynamic and non-progressive settings, this is more challenging, as algorithms must adapt to changing network structures and reversible node states. Moreover, the absence of labeled datasets for the IM problem limits the applicability of supervised learning approaches. Deep reinforcement learning (RL) has proven to be an effective solution to these challenges \cite{arulkumaran2017deep}, as it efficiently learns and optimizes strategies through interaction with the environment, eliminating the need for labeled datasets and showing great promise for CO problems \cite{mazyavkina2021reinforcement}. 

Therefore, in this paper, we propose to model the focused dynamic non-progressive IM problem as a Markov Decision Process \cite{bellman1957markovian} and employ Double Deep Q-network (Double DQN) \cite{van2016deep}, a deep RL algorithm, to learn the optimal policy. To address the challenge posed by the high dimensionality and sparsity of dynamic social networks, we extend a state of the art (SOTA) dynamic graph embedding method called Temporal Graph Networks (TGNs) \cite{rossi2020temporal}. TGNs can well generate low-dimensional dense representations to capture features of each node at each timestamp. In this work, for the focused IM problem, we propose an extension of TGNs to generate an aggregate representation for each node that can capture a long-term evolving history of the node features. 
Furthermore, we propose a novel influence estimator module that can effectively utilise the node representations to estimate the influence of each node and can well incorporate with Double DQN.
To the best of our knowledge, this is the first method for solving the dynamic non-progressive IM problem.

Besides the proposed method, we also focus on diffusion models that are used to describe the influence diffusion in social networks. The most used diffusion model to describe such non-progressive influence diffusion is the Susceptible-Infected-Susceptible (SIS) \cite{kermack1927contribution}. It has several variants and also has been defined on dynamic networks. However, we argue that they exhibit limitations for modeling the influence spread in dynamic social networks as the SIS model is originally designed to describe the epidemic spread. Therefore, we propose a Social-SIS diffusion model, which extends the well-recognized continues-time SIS model \cite{fennell2016limitations} to better capture the non-progressive influence diffusion in dynamic social networks. We conducted extensive experiments on three different types of real-world social network datasets and found that our proposed method outperforms all included benchmark methods. 
% The only exception occurs with small and densely connected networks, where the greedy methods outperform. However, we note that greedy algorithms come with high running time.  

We summarize our contributions as follows: 1) we propose a novel dynamic non-progressive IM approach that integrates dynamic graph embedding with deep RL. Specially, a dynamic graph embedding, TGNs, is extended to generate representations that can capture nodes' evolving history. A novel influence estimator module that efficiently estimate the node influence based on the representations and encapsulates deep RL is also proposed; 2) we extend a well-recognized epidemic model to model the non-progressive influence diffusion in dynamic social networks which is proved by the experiment;
3) we carryout extensive experiments to demonstrate the superiority of the proposed method over SOTA baselines on different types of real-world social networks. \footnote{Upon acceptance of the paper, we will release the code on GitHub.}

%%%%%%%%%%%%%%%%%%%%%%%%%%%%%%%%%%%%%%%%%%%%%%%%%%%%%%%%%%%%%%%%%%%%%%%%

\section{Related Work} 

In this section, we first discuss existing non-progressive diffusion models that are commonly used in IM problems. Then, a general overview of existing solutions to the IM problems is given. 

\subsection{Non-progressive Diffusion Models} \label{RW_DF}
There are two kinds of non-progressive diffusion models in social IM problems \cite{li2018influence}. The most commonly used are Susceptible-Infected-Susceptible (SIS) \cite{kermack1927contribution} and its variants which are originally designed to describe the epidemic spread. In SIS models, nodes have two states: susceptible (inactive) or infected (active). Infected nodes infect (activate) neighbourhood susceptible nodes with probability $\beta$. Infected nodes recover to susceptible state with probability $\mu$ and can be reinfected. Minority studies on non-progressive IM problems also use Voter model \cite{liggett1985interacting} and its variants to describe the influence diffusion. These non-progressive models have been extend to dynamic networks \cite{sun2015contrasting,fernandez2013timing}. The main change is to add time constraints to the diffusion process.

The epidemic model, Susceptible-Infected-Recovered (SIR)\cite{kermack1927contribution}, is also called non-progressive but refers to the epidemic spread. In SIR, infected nodes will turn into the recovered state and cannot infect other nodes nor be infected again. However, Kempe et al. stated that nodes can switch between active and inactive state multiple times under the non-progressive setting in their study \cite{kempe2003maximizing} where first proposed the non-progressive IM. Thus, IM problems using SIR models are \textbf{not} considered non-progressive in this paper.

We argue that neither model is suitable for describing non-progressive influence diffusion in social networks for the following reasons.
The SIS (Susceptible-Infected-Susceptible) model, originally developed to describe the spread of diseases, assumes a fixed recovery probability $\mu$, even when applied to epidemic models on dynamic networks. However, this assumption does not align well with the nature of influence diffusion in social networks. In social networks, disappearance of influence is not a fixed probability; rather, it is often sustained through frequent interactions with advertisers or like-minded individuals \cite{berger1999influence,schmidt2015advertising}. These interactions can reinforce certain beliefs or desires, thereby motivating individuals to continue spreading particular ideas or messages. As a result, the time length that an active node returns to the inactive state should not be a fixed probability. Instead, this should vary depending on the nature and frequency of interactions with other nodes in the network.
%, rather than being governed by a fixed probability.}
Similarly, in the Voter model, a node simply mimics the state of its randomly selected neighbours in order to achieve consensus in the network. On the other hand, IM is concerned with the spread of propagation rather than the consistency of opinions.

\subsection{Solutions to IM Problems}

To the best of our knowledge, existing research in IM has not yet utilized dynamic non-progressive diffusion models to analyze influence diffusion in social networks, i.e., no existing solution to the dynamic non-progressive IM problem. Thus, we give an overview of solutions to other types of IM problems.

\subsubsection{Non-deep learning solutions}
Typical non-deep learning solutions to the IM problems are greedy algorithms that approximate the optimal seed set by iteratively selecting nodes that yield the highest marginal gain in influence spread calculated using Monte Carlo simulation \cite{chen2009efficient, goyal2011celf++, leskovec2007cost,aggarwal2012influential}. Some methods \cite{zhu2023influence,kundu2011new} solve the IM problem by analysing the node importance based on graph statistics such as degree or PageRank. These methods perform well on small scale social networks. 
However, since they do not scale to increasingly large social networks, deep learning-based solutions are becoming mainstream.

\subsubsection{Deep Learning Solutions}
Graph embedding (GE) is used widely in deep learning based IM methods due to the high dimensionality and sparsity of networks. GE densely represents nodes as points in a low-dimensional vector space and reflect the relationships among nodes \cite{cai2018comprehensive}. GE approaches can be broadly classified as static or dynamic. Static GE \cite{grover2016node2vec,ou2016asymmetric} aims to capture the static topological structure and relationships between nodes. In addition to these static features, dynamic GE methods \cite{rossi2020temporal,xu2020inductive} also capture the temporal evolution of the graph.

The FastCover algorithm proposed by Ni et al. \cite{ni2021fastcover} reduces the IM problem to a budget-constrained d-hop dominating set problem (kdDSP) and design a multi-layer GNN which can capture the diffusion process to solve the problem. Zhang et al. \cite{zhang2022network} proposed a self-labeling mechanism and designed a GCN with adjustable number of layers for different sizes of networks to balance scalability and performance. Kumar et al. \cite{kumar2022influence} transfers the IM problem to a pseudo-deep learning regression problem. They use a big real-world dataset to train a Graph Neural Network (GNN) based regressor using the influence of node under SIR and IC model as the labels. 

Deep reinforcement learning (DRL) \cite{li2017deep} offers a new direction for solving the IM problem.
% Reinforcement learning (RL) \cite{kaelbling1996reinforcement} is a machine learning paradigm where an agent learns strategies through trial and error by interacting with an environment, aiming to maximize cumulative rewards. DRL employs deep neural networks to approximate complex decision-making strategies or value functions, enabling agents to effectively learn and make decisions in high-dimensional and complex environments. 
DRL effectively addresses the issue of lacking labeled datasets and excels at solving combinatorial optimization (CO) problems \cite{mazyavkina2021reinforcement}, including IM. 
Li et al. \cite{li2019disco} first use RL to solve the static progressive IM problem and proposed a framework named DISCO, that composes GNN and DQN. Based on DISCO, Wang et al. \cite{wang2021reinforcement} proposed the IMGER. They use graph attention networks for graph embedding. Chen et al. \cite{chen2023touplegdd} proposed ToupleGDD, which is an end-to-end reinforcement learning framework where model is trained on several small randomly generated graphs which makes it not graph-specific. All these methods are designed for static progressive IM problem. 

Some most recent studies \cite{dizaji2024influence, sheng2022dynamic} use deep learning to solve the dynamic progressive IM problem. While they describe a dynamic network as a series of network snapshots which results in information loss.
We describe dynamic social networks using continuous time \footnote{For more details about different representations of dynamic social networks and their comparisons, please refer to Appendix~\ref{appendix_representation_network}.}, allowing a more accurate assessment of node influence.
To the best of our knowledge, there are no deep learning solutions developed for non-progressive IM problems.

%%%%%%%%%%%%%%%%%%%%%%%%%%%%%%%%%%%%%%%%%%%%%%%%%%%%%%%%%%%%%%%%%%%%%%%%

\section{Problem Definition}

We first define the dynamic non-progressive influence maximization problem formally. 

\begin{definition}[Dynamic Non-Progressive IM Problem]
Given a dynamic network $\mathcal{G}$ and a dynamic non-progressive diffusion model, the goal of dynamic non-progressive influence maximization problem is to find a $k$-size subset of nodes whose influence is maximum, i.e: $$\max_{\mathcal{S}\subset \mathcal{V}} I(\mathcal{S})$$ where $|\mathcal{S}|=k$, $\mathcal{V}$ is the node set of $\mathcal{G}$, and $I(\mathcal{S})$ is the influence of the seed set $S$ determined by the diffusion model.
\end{definition}

As discussed in Section~\ref{RW_DF}, existing diffusion models have limitations in accurately capturing non-progressive influence diffusion in dynamic social networks. To address this, we propose a new dynamic non-progressive diffusion model, termed Social-SIS. This model extends the well-recognized continuous-time SIS model~\cite{fennell2016limitations} by introducing a key modification: nodes return to the susceptible (inactive) state based on the influence of their neighboring nodes, rather than following a fixed probability.

Dynamic social networks can be represented using both the discrete and continuous dynamic graph model \cite{xue2022dynamic}\footnotemark[2]. Fennell et al. \cite{fennell2016limitations} proved that the SIS model defined on dynamic networks described using the continuous dynamic graph model can more accurately model the diffusion process. In this paper, we therefore decide to use the continuous model to capture temporal features of the dynamic social networks, so that the influence of the seed set is evaluated more accurately. 

The formal definition of Social-SIS is presented below and detailed in Algorithm~\ref{alg_Social-SIS}.

\begin{algorithm}
\caption{Social-SIS}
\label{alg_Social-SIS}

\textbf{Input}: $\mathcal{G}=(\mathcal{V},\mathcal{E}(t))$, $\mathcal{S}$, $\mu$, $t_{act}$ 
\begin{algorithmic}[1] %[1] enables line numbers

\FOR{$seed \textbf{ in } \mathcal{S}$}
    \STATE $act\_stats[seed]=[T_s, T_e]$
\ENDFOR
\STATE $\mathcal{E} \leftarrow$  Sort edge set $\mathcal{E}(t)$ according to edge timestamp
\FOR{$(v_s, v_e, t) \textbf{ in } \mathcal{E}$} 
    \IF{$v_s$ is active at time t }
        \IF{$\textbf{random}(0,1) < \mu$}
            \STATE $t_{new\_act\_s}=max(t, v_{act\_e}^k)$
            \STATE $t_{new\_act\_e}=min(T_e, t_{new\_act\_s}+t_{act})$
            \STATE Append $(t_{new\_act\_s}, t_{new\_act\_e})$ to $act\_stats[v_e]$
        \ENDIF
    \ENDIF
\ENDFOR
\STATE \textbf{return} $act\_stats$
\end{algorithmic}
\end{algorithm}

\paragraph{Network Definition}A dynamic network existing from $T_s$ to $T_e$ (specific time points) is described using continuous dynamic graph model as $\mathcal{G}=(\mathcal{V}, \mathcal{E}(t))$, where $\mathcal{V}$ is the node set that includes all users appear in this network. The number of nodes is denoted as $N=|\mathcal{V}|$. $\mathcal{E}(t)$ is the edge set where each edge is denoted as $e(v_s,v_e,t)$, where $v_s$ and $v_e$ are the start and end node of edge $e$ respectively. $t$ is the timestamp indicating when the edge is valid. 

\paragraph{Influence Diffusion}All the nodes in the seed set $\mathcal{S}$ are always active. For each edge $(v_s, v_e, t)$ in $\mathcal{E}(t)$, $v_s$ will try to activate $v_e$ at time $t$ if and only if $v_s$ is in the active state at time $t$. Activation is successful with a probability of $\mu$. We modify the continues-time SIS model by introducing $t_{act}$. If $v_e$ is activated successfully, it will stay in the active state for $t_{act}$. Notably, $v_e$ can be activated even if it is active, and such a successful activation will prolong the time it remains active.

\paragraph{Influence Calculation} Following \cite{lou2014modeling}, we introduce how to calculate the influence of a seed set under the Social-SIS.  $act\_stats$ is defined to record the active time intervals for each node. Specifically, for node $v$, $act\_stats(v)=\{(v_{act\_s}^1, v_{act\_e}^1),\cdots,(v_{act\_s}^k, v_{act\_e}^k)\}$ representing the $k$ time intervals when node $v$ is in active state. A node may already be activated when another node tries to activate it. We deal with this by consecutively adding up the active periods. Hence $t_{act}$ starts from $t_{new\_act\_s}=\max(t, v_{act\_e}^k)$ and ends at $t_{new\_act\_e}=\min(T_e, t_{new\_act\_s}+t_{act})$. The influence $I$ of a seed set $\mathcal{S}$ is measured by the average length of time that all nodes are in the active state. Specifically, it is defined as follows:
\begin{equation}
    I(\mathcal{S})=\frac{\sum_{i=0}^N \sum_{j=0}^k {v_i}_{act\_e}^j - {v_i}_{act\_s}^j}{N}
\label{Equation:influence}
\end{equation}

Noteworthy, due to the randomness caused by $\mu$ (Line 9 in Algorithm~\ref{alg_Social-SIS}), the calculation needs to be repeated multiple times.

%%%%%%%%%%%%%%%%%%%%%%%%%%%%%%%%%%%%%%%%%%%%%%%%%%%%%%%%%%%%%%%%%%%%%%%%

\section{Methodology}

\begin{figure}
\centering
\includegraphics[width=0.5\textwidth]{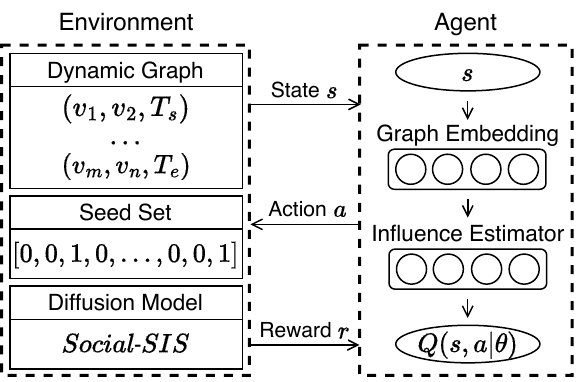}
\caption{Overall framework of the proposed method.
The left side formulates the proposed IM problem as a Markov Decision Process (MDP). The right side uses Double Deep Q-Network (Double DQN) for policy learning and optimization.}
\label{fig:overall_framework}
\end{figure}

In this section, we describe how we solve the proposed dynamic non-progressive IM problem using dynamic graph embedding and deep reinforcement learning. The framework of our method, DNIMRL, is shown in Figure \ref{fig:overall_framework}. The proposed dynamic non-progressive problem is formulated as a Markov Decision Process (MDP) \cite{bellman1957markovian} and the Double DQN method \cite{van2016deep} is employed to learn the optimal policy. 

\subsection{Problem Formulation}
The goal of the defined IM problem is to find a fixed-size subset of users whose influence (measured by a dynamic non-progressive diffusion model) is maximum in the dynamic network. We formulate the problem as a finite and discrete Markov Decision Process (MDP) \cite{bellman1957markovian}. 
The key concepts of this MDP are defined as follows. 

\begin{itemize}
    \item The state $s_t$ at any time step $t$ represents the original dynamic network and the current seed set $\mathcal{S}_t$. The seed set is represented using a one-hot vector. The length of the one-hot vector is equal to the total number of nodes in the graph and each element is a binary indicator signifying whether the corresponding node is in the current seed set.
    \item The action $a_t$ taken at step $t$ refers to adding a node $v$ into the seed set $\mathcal{S}_t$. The node is chosen from the set of nodes that are not in the current seed set, i.e. $v \in \mathcal{V} - \mathcal{S}_t$.
    \item The reward $r_t$ for an action that adds a node $v$ into the seed set $\mathcal{S}_t$ is calculated as the difference in influence before and after adding $v$, i.e. $I(\mathcal{S}_t \cup \{v\})-I(\mathcal{S}_t)$.
\end{itemize}

In this problem, the state includes the dynamic network itself and the current seed set which is empty at the beginning. At each step, the action is defined as adding a node into the seed set. All nodes that are not yet in the seed set can be added. The reward of the action is the marginal gain in the seed set influence from taking this action, which ensures that the cumulative reward from all actions is equivalent to the aggregate influence exerted by the seed set. In this way, learning an optimal policy to guide the selection of actions to maximize the cumulative reward equates to maximizing the total influence propagation across the network, aligning with the objective function of the Influence Maximization Problem in network theory. The process ends when the seed set size reaches the predefined threshold. 

\subsection{Double DQN with Graph Embedding}
We employ the Double DQN \cite{van2016deep} to learn and optimize the policy for selecting seed nodes. Double DQN extends the popular Deep Q-Network (DQN) method, which uses a neural network to approximate the Q-value function (i.e. quantifies the expected future rewards for taking a specific action in a given state). 
Double DQN uses two separate neural networks with same structure: one for selecting actions and another for evaluating those actions. This separation helps mitigate the overestimation of action values often observed in DQN. It enables the model to handle environments with high-dimensional state spaces and leads to more reliable and stable learning outcomes \cite{van2016deep}. 
In this paper, we propose to integrate graph embedding within the learning procedure of Double DQN. 

\subsubsection{Network Architecture} \label{sec:Network Architecture}
\begin{figure}
\centering
\includegraphics[width=0.9\textwidth]{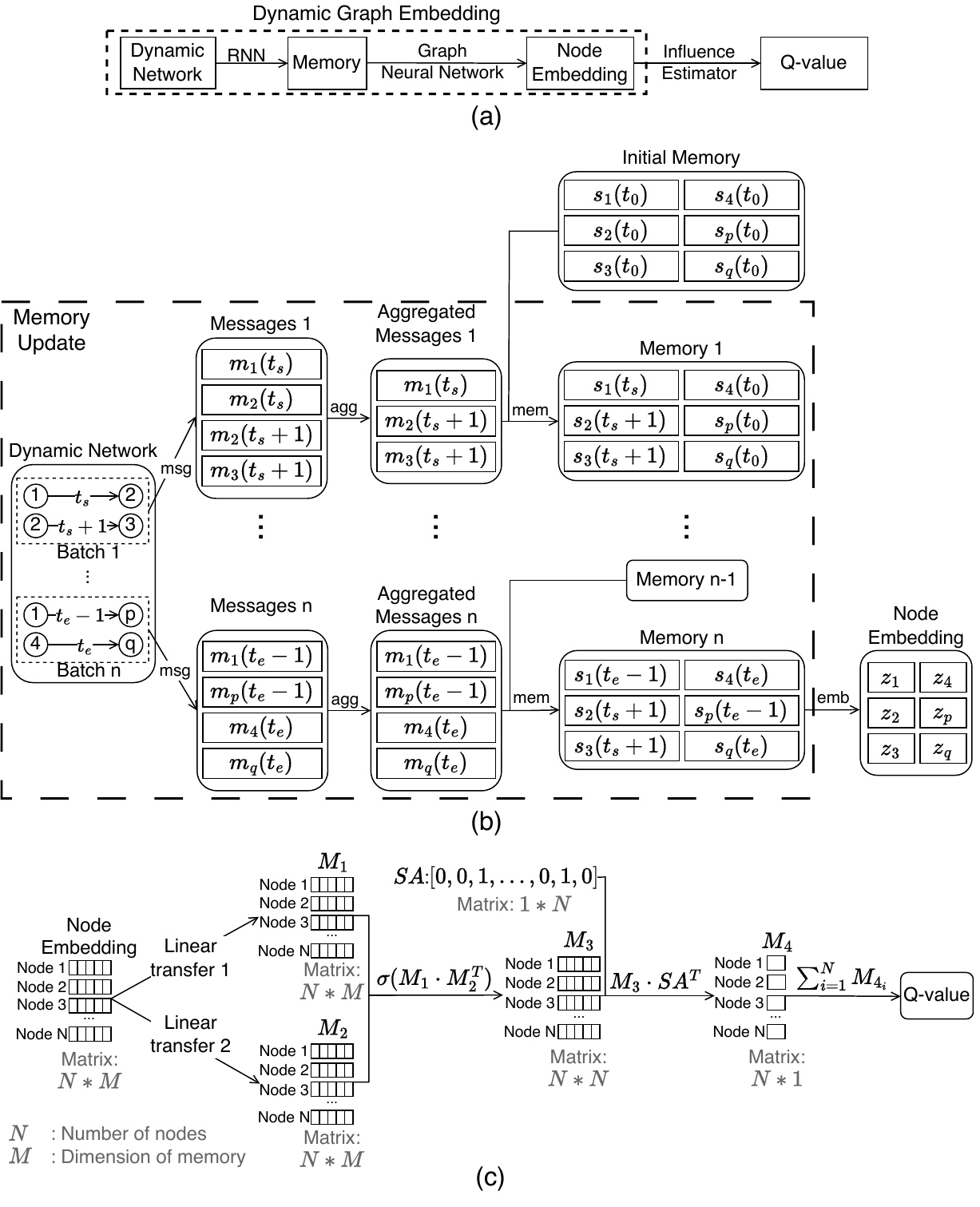}
\caption{Neural network in Double DQN. (a) Architecture overview. (b) Dynamic graph embedding module. The dotted box illustrates how the initial memory is updated. The bottom right shows the generation of node embeddings based on the updated memory. (c) Influence estimator module.}
\label{fig:overall_network_architecture}
\end{figure}

We first introduce the neural network we designed to approximate the Q-value function $Q(s,a)$, which estimates the expected return (total cumulative reward) for taking action $a$ in a particular state $s$. The architecture of the neural network is shown in Figure~\ref{fig:overall_network_architecture}(a). It consists of two primary components: a dynamic graph embedding module and a influence estimator module. The dynamic graph embedding module has two main purposes. The first one is to convert the dynamic network into low-dimensional dense embedding vectors. Second, the embedding vectors are used to capture both the temporal dynamics and structure of the network, serving as the foundational input for the subsequent value calculation process. These embedding vectors are then fed into the influence estimator module designed to approximate the Q-value. 

\paragraph{Dynamic Graph Embedding Module}
This module is designed based on the TGNs proposed by Rossi et al. \cite{rossi2020temporal}, which is a state of the art dynamic graph embedding method. In the original design, TGNs focus on generating embedding vectors for each node at individual time steps. In contrast, we place a greater emphasis on capturing the evolving history of each node. This perspective is more pertinent to the problem of IM, as it allows for a more comprehensive understanding of how nodes influence each other over time. 
%Therefore, we modify the original framework of the TGNs to enhance its ability to form the dynamic graph embedding module (Figure~\ref{fig:overall_network_architecture}(b)). 
Therefore, we modify the original framework of the TGNs to enhance the dynamic graph embedding module (Figure~\ref{fig:overall_network_architecture}(b)). 
Here, we describe our-proposed modifications. For a detailed description of the graph embedding module, please refer to Appendix~\ref{appendix_DGEModule}.

This module consists of two key components: memory and embedding. The memory component is designed to memorize long-term dependencies for each node in the graph. It consists of $|\mathcal{V}|$ vectors, each representing a node, which are initialized as zero vectors. The memory vector of each node is updated upon the generation of an edge related to that node. The set of edges is divided into \(n\) batches, with each batch independently updating the memory. 
We follow the original design of TGNs for the specific update process of a batch.

After updating the memory with all batches, we get the final memory $s'$ which can preliminarily serves to represent the evolving history of nodes. To further integrate information from neighboring nodes and more comprehensively capture the dynamics of the network, graph embedding is conducted. 

Embedding component $emb$ is performed using the temporal graph attention model proposed by TGNs. Different from TGNs that generate embeddings for each node at each timestamp, we only generate one embedding $z_i$ for each node $v_i$ that represent its evolving history due to the different role that generated embedding plays. Thus, we perform graph embedding to the final memory $s'$. The input embedding to the graph attention model of each node $v_i$ is its final memory $s_i'$. For each attention layer $l$, the query $\mathbf{q}^{(l)}$ is defined in the same way as TGNs, the key $\mathbf{K}^{(l)}$ and value $\mathbf{V}^{(l)}$ are still the neighbors of the node $v_i$ but represented using different form as follows:
\begin{equation}
\mathbf{K}^{(l)} = \mathbf{V}^{(l)}= [\mathbf{h}_1^{(l-1)}||\boldsymbol{\phi}(t_1),\ldots,\mathbf{h}_N^{(l-1)}||\boldsymbol{\phi}(t_N)]. 
\end{equation}
where ${\mathbf{h}_1^{(l-1)},\ldots,\mathbf{h}_N^{(l-1)}}$ are the $(l-1)$-th layer representations of these neighborhood nodes, which refers to all nodes $v_1,\ldots,v_N$ that have ever been connected to node $v_i$ through an edge and $t_1,\ldots,t_N$ are the timestamps of these edges. $\boldsymbol{\phi}(\cdot)$ represents a generic time encoding \cite{xu2020inductive} and $||$ is the concatenation operator. The embedding $\mathbf{z}_i$ of node $v_i$ is the output of the last layer, i.e. $\mathbf{z}_i = \mathbf{h}_i^{(L)}$.

\paragraph{Influence Estimator Module}
From the dynamic graph embedding module, we have generated node embeddings that can represent the dynamics of node interconnections within the dynamic network. In the following, we describe how the Q-value function $Q(s,a)$ is approximated based on node embeddings (shown in Figure~\ref{fig:overall_network_architecture}(c)). 

Node embeddings, produced by the dynamic graph embedding module, first undergo two linear transformations to yield two matrices $M_1$ and $M_2$. To perform the linear transformation, node embeddings are processed through two different MLPs. Specifically, for node embedding $\mathbf{z}$: $M_1=\mathrm{MLP_1}(\mathbf{z})$ and $M_2=\mathrm{MLP_2}(\mathbf{z})$. These two linear transformations are designed to extract and enhance the representation of features pertinent to the influence maximization problem from varied perspectives, thereby providing a richer informational foundation for Q-value computation. 

$M_1$ and $M_2$ then undergo a dot product operation to produce a new matrix $M_3$. This operation measures the similarity between node representations. 
%The higher similarity a node pair has indicates the more similar their behaviour is in the network and therefore the more likely they are to influence each other. 
The higher similarity between a node pair indicates that their behavior is more similar in the network, and therefore they are more likely to influence each other.
So, $M_3$ can be used to represent the potential influence of each node on others. To normalize the potential influence of all other nodes on a single node into values between 0 and 1, the sigmoid function ($\sigma$) is applied to each row of $M_3$, i.e. $M_3 = \sigma(M_1 \cdot M_2^T)$. Following this, the state $s$ and action $a$ are represented as a one-hot vector (denoted as $SA$), where nodes already in the seed set and the node selected by the action $a$ are set to 1, with the rest are set to 0. A dot product operation between $M_3$ and $SA$ yields a new Matrix $M= M_3 \cdot {SA}^T$. By doing so, each element in $M$ represents the quantified contribution of each node towards influence maximization under the given action and state. Therefore, summing all elements in $M$ provides the Q-value for the action under the specified state, i.e. $Q(s,a) = \sum_i^{|M|} M_i$.
In summary, the Q-value $Q(s,a)$ is approximated using the node embeddings $\mathbf{z}$ as follows:
\begin{align}
    Q(s,a) &~= \sum_i^{|M|} M_i \\
    M &~=  \sigma(\mathrm{MLP_1}(\mathbf{z}) \cdot \mathrm{MLP_2}(\mathbf{z})^T) \cdot {SA}^T
\end{align} 

\subsubsection{Training Strategy}

In the Double DQN, two neural networks both using the architecture detailed in Section~\ref{sec:Network Architecture}, the Q-network and the target network, are employed. The Q-network $Q(s,a|\theta)$, where $\theta$ is the parameter of the Q-network, directly interacts with the environment, making real-time decisions based on the state inputs. The target network $\hat{Q}(s,a| \hat{\theta})$ provides a stable benchmark for evaluating the decisions made by the Q-network. The target network is updated less frequently by copying the Q-network's parameter after a fixed number of episodes. The loss function is designed to measure the discrepancy between the predicted Q-values by the Q-network and the more stable Q-value targets provided by the target network. The exact equation for the loss function is:
\begin{equation}
    L(\theta) = (r + \gamma \hat{Q}(s_{t+1}, \arg\max_{a'}Q(s_{t+1}, a'|\theta)| \hat{\theta}) - Q(s_t, a_t|\theta))^2
\label{Equ:loss}
\end{equation}
where $r$ is the reward received for taking action $a$ in state $s$, $s'$ is the subsequent state after taking action $a$. $\gamma$ is the discount factor, which balances the importance of immediate versus future rewards.

Two techniques are used in training. First, replay buffer is used to store experience data during interactions between the agent and the environment, allowing these data to be used during training to improve learning efficiency. Second, $\epsilon$-greedy is used to balance exploration and exploitation by allowing random action selection under a small probability. 
%The parameter $\epsilon$ decays over training. 
The specific training strategy and the detailed algorithm are presented in Appendix~\ref{appendix_TS}. 

%%%%%%%%%%%%%%%%%%%%%%%%%%%%%%%%%%%%%%%%%%%%%%%%%%%%%%%%%%%%%%%%%%%%%%%%

\section{Experiments}
% In this section, we demonstrate the feasibility of our method. 
All experiments are conducted on a machine with an Intel Xeon Platinum 8360Y  (2.4 GHz, 18 cores), 128 GiB DDR4 RAM, and a NVIDIA A100 (40 GiB HBM2 memory), running Linux release 8.6.

\subsection{Experiment Setup}
\subsubsection{Dataset}

Three real-world datasets are used. 
% 1) CollegeMsg \cite{panzarasa2009patterns}. This dataset includes users who sent/received messages in a Facebook like Social Network originating from an online community for students at the University of California, Irvine. In the dataset, each node represent a student and each edge represents a message sent from the sender to the receiver. 
1) Bitcoinalpha \cite{kumar2016edge}. Bitcoinalpha is a who-trusts-whom network of people who trade using Bitcoin on a platform called Bitcoin Alpha. In the dataset each node represents a Bitcoin user and each edge represents a credit rating record from the rater to the ratee. 2) Bitcoinotc \cite{kumar2016edge}. This dataset is similar to Bitcoinalpha, but collected from another platform named Bitcoin OTC. 3) Facebook \cite{viswanath2009evolution}. This dataset is collected from New Orleans regional network in Facebook. In the dataset, each node represents a user and each edge represents a friendship between users. 

Bitcoinalpha, and Bitcoinotc are from SNAP\footnote{https://snap.stanford.edu/data/index.html} without modification. Facebook is from NR\footnote{https://networkrepository.com/index.php} and loops (edges that start and end at the same node) are removed. The statistical information of the three used datasets is shown in Table~\ref{table:dataset}. The three datasets were carefully selected so that they increase in size and have different densities (edge to node ratio $|\mathcal{E}| / |\mathcal{V}|$) to measure the applicability of the method on different types of datasets.

\begin{table}
\centering
\caption{Overview of datasets used for the experiments.}
\begin{tabular}{ccccc}
\hline
Dataset      & \shortstack{Nodes \\ $(|\mathcal{V}|)$} & \shortstack{Edges \\ $(|\mathcal{E}|)$} & \shortstack{Density \\ $(|\mathcal{E}|/|\mathcal{V}|)$} & \shortstack{Duration \\ (seconds)} \\ \hline
% CollegeMsg   & 1,899       & 59,835      & 31.5    & 16,736,181         \\
Bitcoinalpha & 3,783       & 24,186      & 6.4     & 164,246,400        \\
Bitcoinotc   & 5,881       & 35,592      & 6.1     & 164,442,412        \\
Facebook     & 45,813      & 855,542     & 18.7    & 134,873,285        \\ \hline
\end{tabular}
\label{table:dataset}
\end{table}

\subsubsection{Baselines}
%Ideally, we want to choose methods designed for dynamic non-progressive IM problem as baselines. However, to the best of our knowledge, there is no existing method designed exactly for this type of IM problem. In this case, for a fair comparison, we chose five methods designed for other types of IM problems.
To the best of our knowledge, currently there are no existing baseline methods tailored for dynamic non-progressive IM problems. Thus, for a fair comparison, we have opted to include five methods designed for addressing other types of IM problems.
The five different methods cover all types of IM problems (see Table~\ref{table:baselines_overview} for an overview of their key features). Among them, CELF~\cite{leskovec2007cost} and INDDSN~\cite{aggarwal2012influential} are the two best-known greedy algorithms designed for static and dynamic IM problems, respectively. 
% CELF can be applied to both progressive and non-progressive IM problems. INDDSN is designed for progressive IM problems only.
These two greedy algorithms are chosen to show the accuracy of our method.
ToupleGDD \cite{chen2023touplegdd} is one of the state-of-the-art (SOTA) methods designed for the static progressive problem, which also employs RL. KTIM \cite{zhu2023influence} is the only SOTA method for the dynamic progressive IM problem which also defines dynamic networks using the continuous model. For non-progressive methods, we chose the SOTA method TSGC \cite{devi2023identification} from limited number of methods.

To make these static methods applicable for the experiments, we transform the dynamic networks into static networks by removing the timestamps of edges and suppressing duplicate edges.
%To make these static methods feasible for the experiments, we transform the dynamic networks into static networks by removing the timestamps of the edges and suppressing duplicate edges and make these static methods work on the static networks. 
%For progressive methods, they can be applied directly as they are designed to use the topology information of the network only. 
The progressive methods can be applied directly (on dynamic networks), as they use the topology of the network only. 

\begin{table}
\caption{Baselines overview. The static / dynamic and progressive / non-progressive indicates the baseline method is designed for static / dynamic and progressive / non-progressive IM problem respectively.}
\centering
\begin{tabular}{cccc}
\hline
Baseline  & \begin{tabular}[c]{@{}c@{}}Static/\\ Dynamic\end{tabular} & \begin{tabular}[c]{@{}c@{}}Progressive/\\ Non-progressive\end{tabular} & Key feature \\ \hline
CELF      & Static         & Both               & Greedy  \\
INDDSN    & Dynamic        & Progressive        & Greedy  \\
ToupleGDD & Static         & Progressive        & SOTA \& RL      \\
KTIM      & Dynamic        & Progressive        & SOTA \& Continuous \\ 
TSGC      & Static         & Non-progressive    & SOTA  \\ \hline
\end{tabular}
\label{table:baselines_overview}
\end{table}

\subsubsection{Hyperparameter Setting}
For the diffusion model, the activation success probability $\mu$ is set to 0.5 (i.e. the probability of each activation to be random) as the four datasets we used are anonymized, which provides no information about the users. 
As there is randomness in the calculation of the seed set influence, the calculation of the influence will be repeated 2,000 times and take the average for a fair comparison. The hyperparameters for baselines are fine-tuned on each dataset according to original papers.

For our method, we first give the parameters related to dynamic graph embedding module. The dimension of both node memory and embedding is 64. The number of graph attention layers is 1 and the number of heads used in each attention layer is 2. The batch size for updating memory is 200. These hyperparameters follow those provided by TGNs~\cite{rossi2020temporal}, except for the dimension of node memory and embedding, which are fine-tuned by ourselves. For the RL part, we fine-tuned the following hyperparameters: the batch size of transactions for updating Q-network is 16, the Q-network is updated every 1 episode. The target network is updated every 20 episodes. The value of $\epsilon$ for epsilon-greedy starts at 1 and decays to 0.2, decreasing by 0.98 every episode. The $\gamma$ in loss function is set to 0.95, following the standard practice. The learning rate, which is shared by the two modules, is optimized to $0.001$.

\subsection{Quantitative \& Scalability Analysis}

\begin{figure}
\centering
\includegraphics[width=0.8\textwidth]{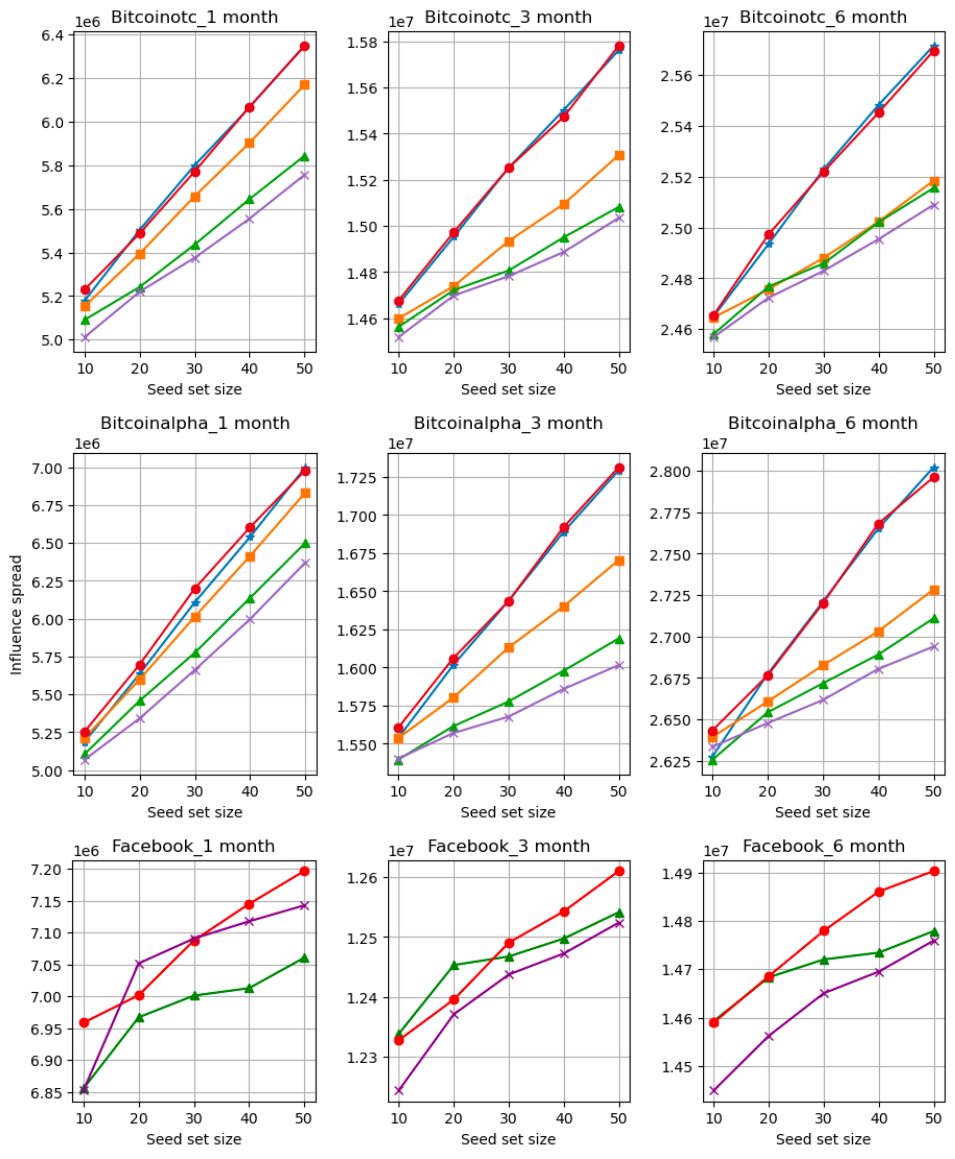}
\caption{Influence of seed sets selected by the evaluated methods under different settings. The title of each sub-graph indicates both the used dataset and the corresponding $t_{act}$ setting in the diffusion model. 
% Since TSGC performed significantly worse than all other methods, the experiment results of TSGC are not shown for the sake of visualisation clarity.
}
\label{fig:seed_set_influence}
\end{figure}

In order to compare the performance of the proposed method with baselines, the influence of the seed set selected by these methods is evaluated. The influence of a seed set is calculated using Equation~\ref{Equation:influence}. The higher the influence of the selected seed set, the higher the performance of the method. As the activation interval $t_{act}$ is an important parameter in our diffusion model, we also consider multiple settings for each dataset. 
% For CollegeMsg, $t_{act}$ is set to 1 day, 7 days and 1 month respectively. 
For all datasets, $t_{act}$ is set to 1 month, 3 months and 6 months respectively. The three increasingly large values of $t_{act}$ simulate the short to long duration of information influence in the real world. In order to make a fair comparison, all experiments were repeated three times and the results were averaged. The experimental results are shown in Figure~\ref{fig:seed_set_influence} and the running time of experiments are shown in Table~\ref{Table:running_time}.

Our method, DNIMRL, is first compared with two greedy algorithms. On the sparse small-scale datasets, Bitcoinalpha and Bitcoinotc, our approach generally outperforms the static greedy method, CELF. When compared to the dynamic greedy algorithm, INDDSN, our method exhibits similar or even superior performance in most cases. Although INDDSN performs slightly better than our proposed DNIMRL on small-scale datasets, the running time of INDDSN is significantly higher. 
% On the denser small-scale dataset, CollegeMsg, our method underperforms relative to CELF. It is noteworthy that the performance of INDDSN is similarly poor on this dataset. This may be attributed to the density of the dataset, which diminishes its dynamic nature, making it more akin to a static graph. 
Importantly, both greedy algorithms are unable to process the large-scale Facebook dataset within 24 hours, whereas our method continues to demonstrate good performance. 

In subsequent comparisons with three of the latest methods, the results of TSGC are not shown in Figure~\ref{fig:seed_set_influence} nor compared due to its significantly lower performance. 
% For example, the influence of the seed set ($\mathcal{|S|}=10$, $t\_act=$1 day) selected by TSGC in CollegeMsg is 88,131 which is an order of magnitude lower than that selected by other methods. 
Compared to the other two methods, ToupleGDD and KTIM, our approach exhibits superior performance on the Bitcoinalpha and Bitcoinotc datasets. 
% However, on the CollegeMsg dataset, our method is outperformed by these competitors. 
On the Facebook dataset, our method holds an advantage, particularly when $t_{act}$ is large or the seed set size is large. Additionally, ToupleGDD can train a generic model for all datasets, but our method needs to be trained for each dataset. However, the training time (32 h) is long and it performs worse than ours. This aspect potentially compensates for the need to repetitively train our model for each dataset. Moreover, training our model specifically for each dataset significantly enhances the quality of the seed set.

In summary, these experiments show that our proposed method demonstrates superiority 
% on sparser datasets when 
compared with state of the art methods even with greedy algorithms. In addition, it has been shown to scale well to large datasets. 

\begin{table}
\centering
\caption{Running time overview (when the size of a seed set is 10).}
\begin{tabular}{ccccc}
\hline
\multicolumn{2}{c}{}                                                                                                               & Bitcoinalpha     & Bitcoinotc          & Facebook          \\ \hline
\multicolumn{2}{c|}{CELF}                                                                                                          & 15 min           & 33 min              & \textgreater 24 h \\ \hline
\multicolumn{2}{c|}{INDDSN}                                                                                                        & 52 min           & 2.3 h               & \textgreater 24 h \\ \hline
\multicolumn{1}{c|}{\multirow{2}{*}{ToupleGDD}}                                                   & \multicolumn{1}{c|}{Training}  & \multicolumn{3}{c}{32h}                                    \\ \cline{2-5} 
\multicolumn{1}{c|}{}                                                                             & \multicolumn{1}{c|}{Execution} & 1.2 sec          & 1.2 sec             & 3.8 sec           \\ \hline
\multicolumn{2}{c|}{KTIM}                                                                                                          & 0.7 sec          & 1.5 sec             & 8 min             \\ \hline
\multicolumn{2}{c|}{TSGC}                                                                                                          & 4 min            & 10 min              & 12.8 h            \\ \hline
\multicolumn{1}{c|}{\multirow{2}{*}{\begin{tabular}[c]{@{}c@{}}DNIMRL\\ (Proposed)\end{tabular}}} & \multicolumn{1}{c|}{Training}  & 20 min           & 26 min              & 13.4 h            \\ \cline{2-5} 
\multicolumn{1}{c|}{}                                                                             & \multicolumn{1}{c|}{Execution} & 4 sec / 10 nodes & 14.4 sec / 10 nodes & 6 min / 10 nodes  \\ \hline
\end{tabular}%
\label{Table:running_time}
\end{table}

\subsection{Ablation Study}
In our Double DQN method, we performed two linear transformations to the node embedding to extract and enhance the representation of influence maximization related features, i.e. the generation of $M_1$ and $M_2$ in Figure~\ref{fig:overall_network_architecture}(c). 
%In order to prove the importance of this step as it consumes extra time, we compare the performance of method using the proposed network architecture and reduced network architecture. 
To show the importance of this step, given that it requires considerable time, we compare the performance of the method using the proposed network architecture with that using the reduced network architecture.
In the reduced network architecture, the two linear transformations are removed and instead using the embeddings to calculate $M_3$ directly, i.e. $M_3=\sigma(\mathbf{z}\cdot \mathbf{z}^T)$. 

As the results from Table~\ref{Table:ablation} demonstrate, the influence of the seed sets selected using the reduced network has a clear reduction compared to the seed sets selected using the proposed network. However, the increase of the running time for each episode is only 0.53 second (from 10.71 s to 10.18 s). The results indicate that this design is realised without a significant increase in running time.

\begin{table}
\centering
\caption{Ablation study on Bitcoinalpha dataset. The influence decrease indicates the decrease on the influence of seed sets selected with and without two linear transformations.}
\begin{tabular}{cccccc}
\hline
Seed Set Size$|\mathcal{S}|$ & 10 & 20 & 30 & 40 & 50 \\ \hline
\begin{tabular}[c]{@{}c@{}}Influence Decrease\end{tabular} & 0.52\% & 0.82\% & 0.48\% & 0.23\% & 0.41\% \\ \hline
\end{tabular}%
\label{Table:ablation}
\end{table}

\subsection{Qualitative Analysis}
\begin{figure}
\centering
\includegraphics[width=\columnwidth]{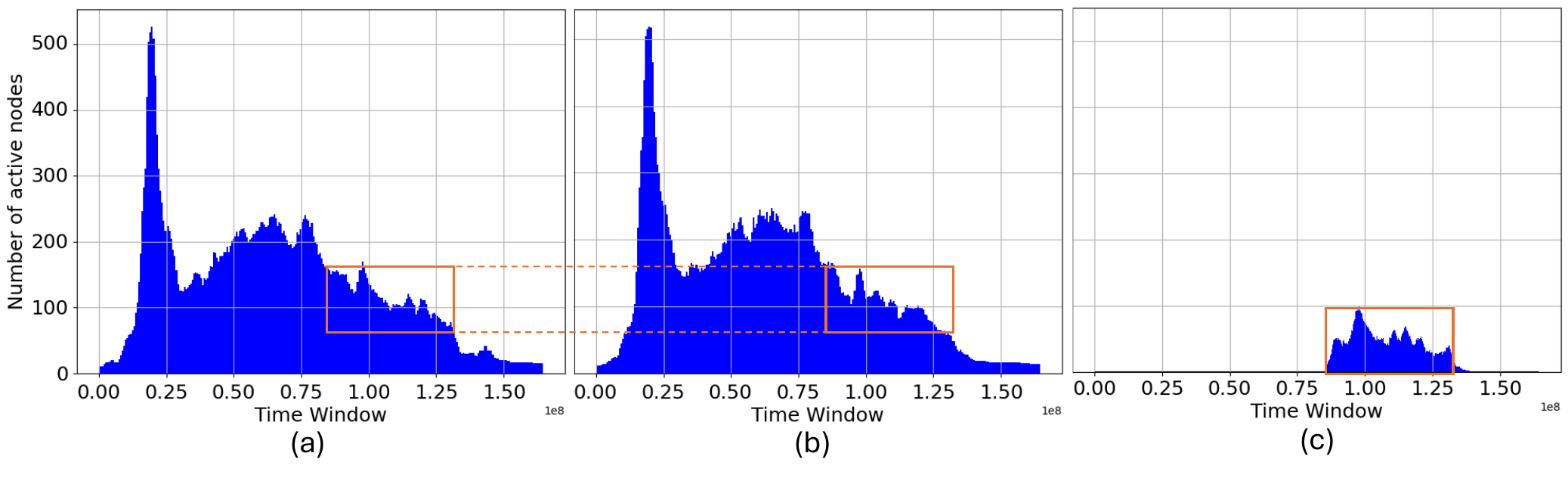}
\caption{Number of active nodes in time windows. (a)-(b): activated with seeds selected by DNIMRL and CELF, (c) activated with seed 3149.}
\label{fig:qualitative}
\end{figure}

%We first further illustrate the validity of the DNIMRL method using an exact example.
We further illustrate the validity of the DNIMRL method.
In Bitcoinalpha, we identify and compare the seed sets chosen by the DNIMRL and CELF. The seed set size is fixed at 10. In Figure~\ref{fig:qualitative}(a)\&(b), we visualise the number of active nodes in different time windows.

First, it can be seen that the number of nodes activated by CELF seeds is significantly lower than the number of nodes activated by DNIMRL seeds in the time windows shown in the orange box. This also contributed to the slightly lower performance of CELF compared to DNIMRL. We analyse the causes as follows.

Here are the seed sets chosen by the two algorithms respectively. DNIMRL: \{3149, 892, 2388, 14, 10, 1536, 246, 18, 20, 1423\}, CELF: \{7, 0, 27, 6, 2, 17, 40, 64, 9, 12\}. The smaller the node's index, the earlier an edge is connected to it in the network. It can be seen that the greedy algorithm tends to select nodes that appear early because such nodes have better connectivity with other nodes. However, the nodes that these seeds can connect to may have a high degree of overlap and may only have a short-term influence when nodes can return to an inactive state.
In contrast, we found an interesting seed node selected by DNIMRL, No. 3149, which appears quite late in the dynamic network. We further analyse all the nodes activated by No. 3149 and visualize the number of these activated nodes over time in Figure~\ref{fig:qualitative} (c).
It can be seen that although the active time of this seed node is limited, its influence period falls within the orange box of \ref{fig:qualitative}(a)\&(b). Therefore, we believe that selecting this node as seed contributed significantly to DNIMRL's excellent performance during this period.
We also calculated the centrality of seed No. 3149, which is lower than other seed nodes. Still, it cooperates with other seed nodes and enhances the overall influence by activating many nodes in this time period. This demonstrates that deep RL algorithms can fully explore the action space and trade-offs between different actions to select the best seed set, which leads to the superiority of our method.

In Figure~\ref{fig:qualitative}(a)\&(b), it can also be seen that the number of active nodes spiked in earlier time windows but then dropped sharply. It indicates information diffusion is very active in the early stage of this dynamic network, and its influence is shrinking afterwards. On the other hand, since the number of active nodes does not drop over time in the progressive setting, such a dynamic phenomenon is not able to be captured. This example further illustrates the importance of studying the dynamic non-progressive setting for IM.

\subsection{Validation of Social-SIS}
As discussed in Section~\ref{RW_DF}, to align with the IM properties of dynamic social networks, the proposed Social-SIS model aims to make the duration for which an active node remains active dependent on its interactions with other nodes, rather than relying solely on a fixed probability.
To experimentally demonstrate this, in Table~\ref{table_paa}, we calculate and present the percentage of successfully activated nodes that were already active.
The results show that at least 50\% of successful activations are performed on already activated nodes. 
This implies that a significant number of activated nodes are influenced by interactions with active neighboring nodes. It shows Social-SIS's ability to capture fluent interactions among nodes.

\begin{table}
\caption{Percentage of activation performed on active nodes. For all datasets, the seed set size is 10 and $t_{act}$ is one month. }
\begin{tabular}{ccccc}
\hline
Dataset    & Bitcoinalpha & Bitcoinotc & Facebook \\ \hline
Percentage & 52.90\%      & 57.32\%    & 84.09\%  \\ \hline
\end{tabular}%
\label{table_paa}
\end{table}

%%%%%%%%%%%%%%%%%%%%%%%%%%%%%%%%%%%%%%%%%%%%%%%%%%%%%%%%%%%%%%%%%%%%%%%%

\section{Conclusion}
In this paper, we investigate the dynamic non-progressive IM problem. We model the problem as a Markov Decision Process and employ a deep reinforcement learning algorithm to find the optimal seed set. In particular, we propose a novel node influence estimation module to validate the temporal influence of nodes based on dynamic graph embeddings. We also extend a dynamic non-progressive epidemic model to capture the influence diffusion in dynamic social networks.
The experiments show that our proposed method works well on different types of social networks 
% (except densely connected small-scale dynamic networks)
and has the ability to scale to large dynamic networks. 
% Improving the performance on densely-connected networks will be a subject of future work. 
It would also be interesting to find new training strategies to obtain a model that can be applied to multiple social networks.

%%%%%%%%%%%%%%%%%%%%%%%%%%%%%%%%%%%%%%%%%%%%%%%%%%%%%%%%%%%%%%%%%%%%%%%%

\bibliographystyle{ACM-Reference-Format}
\bibliography{references}

\newpage

\appendix

\section{Discrete and continuous dynamic graph models} \label{appendix_representation_network}

In this section, we illustrate the discrete and continuous dynamic graph models used to describe dynamic networks which are defined based on the models \cite{xue2022dynamic} commonly used in the AI avenue. Graph is a data structure composed of nodes (vertices) and edges (links) used to model entities and their relationships. Dynamic graph is a special type of graph whose structure (nodes or edges) changes over time. 

In the continuous dynamic graph model, a dynamic network is modeled as a sequence of interactions (edges) between nodes. These interactions and specific occurrence time (usually in second) are recorded. Specifically, we define $\mathcal{G}=(\mathcal{V}, \mathcal{E}(t))$, where $\mathcal{G}$ is a dynamic graph representing a dynamic network exists from time $T_s$ to time $T_e$ (both $T_s$ and $T_e$ are specific time points). $\mathcal{V}$ is the node set that includes all users that have appeared in this network. The number of nodes is denoted as $N=|\mathcal{V}|$. $\mathcal{E}(t)$ is the edge set where each edge is denoted as $e(v_s,v_e,t)$, where $v_s$ and $v_e$ are the start and end node of edge $e$ respectively. $t$ is the timestamp indicating when the edge is valid.

In the discrete dynamic model model, the existence time (from time $T_s$ to time $T_e$) of a dynamic network is divided into $n$ intervals (length usually a day or a week) $\{TI_1,..., TI_k, ..., TI_n\}$, where $TI_k$ starts form $(k-1)*\frac{T_e-T_s}{n}$ and ends at $k*\frac{T_e-T_s}{n}$. All the users and interactions appeared in one interval $TI_k$ form a static graph $\mathcal{G}_k$ (snapshot of the dynamic network). That is, the dynamic network is represented by a sequence of static graphs $ \mathcal{G} = \{\mathcal{G}_1,..., \mathcal{G}_k, ..., \mathcal{G}_n\}$.

Compared with the discrete model, the continuous model provides a higher time resolution. This detailed time recording allows the model to capture more accurately the dynamic changes of the network over short periods of time.

\section{Dynamic Graph Embedding Module} \label{appendix_DGEModule}
This module is designed based on the TGNs proposed by Rossi et al. \cite{rossi2020temporal}, which is a state of the art dynamic graph embedding method. In the original design, TGNs focus on generating representations for each node at individual time steps. In contrast, we place a greater emphasis on capturing the evolving history of each node. This perspective is more pertinent to the problem of IM, as it allows for a more comprehensive understanding of how nodes influence each other over time. Therefore, we extend the original framework of the TGNs to enhance its ability to form the dynamic graph embedding module (Figure~\ref{fig:overall_network_architecture}(b)) which consists of two key components: memory and embedding.

The memory component is designed to memorize long term dependencies for each node in the graph. It consists of $|\mathcal{V}|$ vectors for each node, which are initialized as zero vectors. The memory vector of each node is updated upon the generation of an edge related to that node. The set of edges is divided into \(n\) batches, with each batch independently updating the memory. For a batch, the specific update process is as follows:

\begin{itemize}
    \item \textbf{Message Function $msg$:} For each edge involving node $i$, a message is computed to update its memory. For an edge $(v_s, v_e, t)$, two messages will be generated: $m_{v_s}(t) =$ $ s_{v_s}(t^-)||s_{v_e}(t^-)||\Delta t$ and $m_{v_e}(t) = s_{v_e}(t^-)||s_{v_s}(t^-)||\Delta t$, where $m_{v_s}(t)$ is the message generated for node $v_s$ at time $t$, $s_{v_s}(t^-)$ is the latest memory of node $v_s$ before time $t$, $\Delta t = t-t^-$, and $||$ represents the concatenation operation between two vectors. 

    \item \textbf{Message Aggregator $agg$:} Due to the large number of edges, there may be multiple messages about node $v$ within a batch. For efficiency reasons, multiple messages for a node in a batch $[m_v(t_1), m_v(t_2),...,m_v(t_b)]$ ($t_1<t_2<...<t_b$) are aggregated by remaining most recent messages, i.e. $m_v(t_b)$.

    \item \textbf{Memory Updater $mem$:} The memory vector of a node is updated upon each message involving the node itself using GRU \cite{cho2014learning}. Specifically, $s_v(t) = GRU(m_v(t), s_v(t^-))$.
\end{itemize}

Upon processing all batches, we obtain the final memory for all nodes. At this stage, the memory can preliminarily serve to represent the evolving history of nodes. However, to further integrate information from neighboring nodes and more comprehensively capture the dynamics of the network, graph embedding is conducted. 

Embedding component $emb$ that performs graph embedding based on the memory serves two main purposes. The first is that it allows the further representation of the interactions between nodes and the evolution of their connection patterns. Another role is to address so-called memory staleness problem \cite{kazemi2020representation} which is mentioned in the original paper introducing TGNs. In general, this problem refers to the fact that some users may not be involved in the network for a long time, which can make their memory unable to represent their behavior efficiently. The graph embedding is performed based on the temporal graph attention model proposed in the original TGNs paper. The temporal graph attention model is formed by a series of L graph attention layers, which generate a node's embedding by aggregating information from its L-hop temporal neighborhood. The structure of $l$-th layer is as follows:
\begin{align}
\mathbf{h}_i^{(0)}& =s'_i, \\
\mathbf{h}_i^{(l)}& =\mathrm{MLP}^{(l)}(\mathbf{h}_i^{(l-1)}||\tilde{\mathbf{h}}_i^{(l)}), \\
\tilde{\mathbf{h}}_i^{(l)}& =\mathrm{MultiHeadAttention}^{(l)}(\mathbf{q}^{(l)},\mathbf{K}^{(l)},\mathbf{V}^{(l)}), \\
\mathbf{q}^{(l)}& =\mathbf{h}_i^{(l-1)}||\boldsymbol{\phi}(0), \\
\mathbf{K}^{(l)}& = \mathbf{V}^{(l)}=\mathbf{C}^{(l)}, \\
\mathbf{C}^{(l)}&= [\mathbf{h}_1^{(l-1)}||\boldsymbol{\phi}(t_1),\ldots,\mathbf{h}_N^{(l-1)}||\boldsymbol{\phi}(t_N)]. 
\end{align}
where $s'_i$ is the final memory vector of node $i$, $\mathbf{h}_i^{(l)}$ is the $l$-th layer representation of node $i$. All nodes $v_1,\ldots,v_N$ that have ever been connected to node $v$ through an edge are the temporal neighborhoods of node $v$. $t_1,\ldots,t_N$ are the timestamps of these edges and ${\mathbf{h}_1^{(l-1)},\ldots,\mathbf{h}_N^{(l-1)}}$ are the $(l-1)$-th layer representations of these neighborhood nodes. $\mathbf{q}$, $\mathbf{K}$, $\mathbf{V}$ are query, key and value in attention mechanism respectively.
$\mathrm{MLP}$ refers to a Multilayer Perceptron and $\mathrm{MultiHeadAttention}$ pertains to the multi-head attention mechanism, a pivotal concept introduced by Vaswani et al. \cite{vaswani2017attention}. $\boldsymbol{\phi}(\cdot)$ represents a generic time encoding \cite{xu2020inductive}. The embedding $\mathbf{z}_i$ of node $v_i$ is the output of the last layer, i.e. $\mathbf{z}_i = \mathbf{h}_i^{(L)}$.

\section{Training Strategy} \label{appendix_TS}

In following, we illustrate our training strategy based on Double DQN (cf. Algorithm~\ref{alg:training}). We make use of the two technologies:
\begin{itemize}
    \item \textbf{Replay buffer} collects and stores the agent's experiences at each time step, defined by tuples of states, actions, rewards, and subsequent states, i.e. $(s, a_t, r_t, s_{t+1})$. Training with mini-batches of experiences randomly sampled from this buffer helps to decorrelate the experiences and mitigate the risks of overfitting and instability in learning.
    \item \textbf{Epsilon greedy} manages the trade-off between exploration of the environment and exploitation of known rewarding actions. Initially, a higher $\epsilon$ value prompts the agent to explore broadly, gaining varied experiences within the environment. Over time, $\epsilon$ decays according to a predefined schedule, gradually shifting the agent's strategy towards exploiting its accumulated knowledge to maximize rewards. 
\end{itemize}

First, the exploration rate $\epsilon$ and the replay buffer $D$ are initialized (Line 2-3). The Q network and target network are initialized using the neural network we introduced in Section 4.2.1 of the main paper with the same random weights (Line 4-5). 

For each episode (Line 6-23), the state is first reset to a starting state (Line 7). Within each episode, the agent iterates through a series of actions, with the number of actions capped at a predefined seed set size (Line 8-15). The actions is selected using epsilon greedy. Controlled by $\epsilon$, actions are chosen randomly or selected based on the highest Q-value predicted by the Q network (Line 9). For every action executed, the agent observes the outcome in terms of the reward and the subsequent state (Line 10). 

The transaction $(s, a_t, r_t, s_{t+1})$ whose reward is larger than a certain threshold $\theta_t$, will be stored in the replay buffer $D$ (Line 11-13). The reason for this is that there may be a large number of nodes in the network with very poor connectivity to other nodes, which will make the rewards for actions that select them into the seed set very low. Excluding them from the replay buffer can significantly enhance learning efficiency, as it reduces noise in the learning process and focuses attention on more valuable experiences. By prioritizing transactions that offer higher immediate rewards, this method accelerates the learning process, enabling the algorithm to converge more rapidly to high-quality strategies. Additionally, it prevents the learning process from overfitting to frequent but low-reward transactions, thereby improving the overall quality of the strategy. 

The current state will be updated after an action is executed (Line 14). After the selection of a series of actions options has completed, a mini-batch of experiences is sampled from the replay buffer (Line 16). This batch is then used to update the Q network (Line 17-18) according to Equation~\ref{Equ:loss}. The exploration rate $\epsilon$ will decay $\epsilon_{\text{decay}}$ every episode until it reaches the minimum exploration rate $\epsilon_{\text{min}}$ (Line 19). The target network will be updated using the the parameters of Q network every $T$ episodes (Line 20-22).  

\begin{algorithm}
\caption{Training strategy}
\textbf{Input:} Number of episodes $M$, seed set size $k$, discount factor $\gamma$, learning rate $\alpha$, initial exploration rate $\epsilon_{\text{initial}}$, minimum exploration rate $\epsilon_{\text{min}}$, exploration decay rate $\epsilon_{\text{decay}}$, size of replay buffer $B$, batch size $b$, target network update frequency $T$, transaction reward threshold $\theta_t$
\begin{algorithmic}[1]
    \STATE Initialize exploration rate $\epsilon=\epsilon_{\text{initial}}$
    \STATE Initialize replay buffer $D$ with size $B$
    \STATE Initialize Q network $Q(s,a|\theta)$ with random weights
    \STATE Initialize target network $\hat{Q}(s,a|\hat{\theta})$ with $\hat{\theta}=\theta$
    \FOR{episode = 1 \textbf{to} $M$}
        \STATE Reset environment state $s$
        \FOR{t = 1 to $k$}
            \STATE $a_t =
                    \begin{cases} 
                        \textbf{random}(a), & \text{if } \textbf{random}(0,1) < \epsilon \\
                        \arg\max_a Q(s, a| \theta), & \text{otherwise}
                    \end{cases}$
            \STATE Calculate reward $r_t$ and next state $s_{t+1}$
            \IF{$r_t > \theta_t$}
                \STATE Store transition $(s, a_t, r_t, s_{t+1})$ in replay buffer $D$
            \ENDIF
            \STATE Set $s = s_{t+1}$
        \ENDFOR
        \STATE $mini\_batch$ = Sample $b$ transitions from $D$
        \STATE Calculate loss $L(\theta)$ using $mini\_batch$ based on Equation~\ref{Equ:loss} 
        \STATE Perform a gradient descent step to update Q network
        \STATE Update $\epsilon = \max(\epsilon_{\text{min}}, \epsilon * \epsilon_{\text{decay}})$
        \IF{episode $\% T==0$}
            \STATE Update target network: $\hat{\theta} = \theta$
        \ENDIF
    \ENDFOR
\end{algorithmic}
\label{alg:training}
\end{algorithm}

\end{document}